\documentclass[aps,
footinbib,
superscriptaddress,
showpacs,twocolumn,
prl]{revtex4-1}

\newcommand{\expec}[1]{\langle #1\rangle}

\usepackage{amsmath}
\usepackage{easybmat}
\usepackage{amsfonts}  
\usepackage{enumitem}
\usepackage{graphicx}
\usepackage{amsthm}
\usepackage{subfigure}
\usepackage{array}
\usepackage{hyperref}
\usepackage[usenames]{color}
\usepackage{centernot}
\usepackage{float}

\usepackage{rotating}

\usepackage{braket}

\usepackage{printlen}

\renewcommand{\v}[1]{\mathbf{#1}}
\renewcommand{\vec}[1]{\v{#1}}

\newcommand{\beq}{\begin{eqnarray*}}
\newcommand{\eeq}{\end{eqnarray*}}

\newcommand{\be}{\begin{eqnarray}}
\newcommand{\ee}{\end{eqnarray}}

\newcommand{\assignment}[1]{}

\def\lsim{\mathrel{\rlap{\lower4pt\hbox{\hskip1pt$\sim$}}
    \raise1pt\hbox{$<$}}}                

\def\gsim{\mathrel{\rlap{\lower4pt\hbox{\hskip1pt$\sim$}}
    \raise1pt\hbox{$>$}}}                

\begin{document}

\title{Quantum dynamics from a numerical linked cluster expansion} 
\author{Ian G. White} 
\email{igw@tamu.edu}
\affiliation{Department of Physics and Astronomy, Rice University, Houston, Texas 77005, USA}
\affiliation{Rice Center for Quantum Materials, Rice University, Houston, Texas 77005, USA}
\affiliation{Department of Physics and Astronomy, Texas A\&M University, College Station, Texas 77843, USA}
\author{Bhuvanesh Sundar} 
\email{bs55@rice.edu}
\affiliation{Department of Physics and Astronomy, Rice University, Houston, Texas 77005, USA}
\affiliation{Rice Center for Quantum Materials, Rice University, Houston, Texas 77005, USA}
\author{Kaden R.~A. Hazzard} \email{kaden.hazzard@gmail.com}
\affiliation{Department of Physics and Astronomy, Rice University, Houston, Texas 77005, USA}
\affiliation{Rice Center for Quantum Materials, Rice University, Houston, Texas 77005, USA}

\begin{abstract}
We demonstrate that a numerical linked cluster expansion method is a powerful tool to calculate quantum dynamics. We calculate the dynamics of the magnetization and spin correlations in 
the two-dimensional 
transverse field Ising and XXZ models evolved from a product state. Such dynamics are directly probed in ongoing experiments in ultracold atoms, molecules, and ions. We show that a numerical linked cluster expansion gives dramatically more accurate results at short-to-moderate times than exact diagonalization, and simultaneously requires fewer computational resources. 
More specifically, the cluster expansion frequently produces more accurate results than an exact diagonalization calculation that would require $10^{5}$--$10^{10}$ more computational operations and memory. 
\end{abstract} 

\maketitle

\paragraph{Introduction.} The dynamics of quantum matter is linked to profound questions  across physics. 
How can a system thermalize? 
When it fails to thermalize, what laws replace usual statistical mechanics and thermodynamics? When is dynamics universal?
What novel correlations and phases of matter exist out of equilibrium?
These questions are ripe for progress, 
especially due to the unprecedented control achieved in 
modern experiments in ultracold matter~\cite{polkovnikov:nonequilibrium_2011,lamacraft-moore:potential-insights_2012,langen:ultracold_2015,
altman:nonequilibrium_2015} and solid state systems~\cite{basov:electrodynamics_2011,nicoletti:nonlinear_2016,
gianetti:ultrafast_2016}. 

However, these experiments are rapidly outpacing theory. 
Ultimately, this is due to the fundamental difficulty posed by quantum mechanics: the exponential growth of the Hilbert space with system size. 
Some analytic~
\cite{bray:theory_1994,calabrese:ageing_2005,henkel:non-equilibrium_2010,henkel:non-equilibrium_2009,kamenev:field_2011,taeuber:critical_2014}  and numerical methods work well in special cases. Notable examples are semiclassical methods~
\cite{walls-milburn_1994,PhysRevA.68.053604,blakie:dynamics_2008,gardiner-zoller_2004,POLKOVNIKOV20101790,PhysRevX.5.011022,1367-2630-17-6-065009,PhysRevB.93.174302,orioli:nonequilibrium_2017} and tensor network methods in one dimension~\cite{schollwoeck:density-matrix_2011}.
Nevertheless, exact diagonalization (ED) is often the only applicable tool, and it is limited to very small systems, just two or three sites wide in three-dimensional lattices~\cite{sandvik:computational_2010}.  
Thus new computational methods are urgently needed to understand and drive  experiments, and to tackle the questions posed above. 

\begin{figure}[t]\centering
\includegraphics[width=.95\columnwidth]{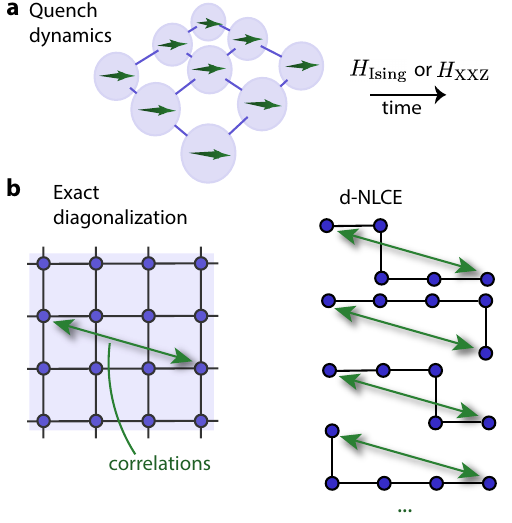}
\caption{(Color online)
Dynamics considered in this paper and intuition behind the dynamical numerical linked cluster expansion (d-NLCE) method. (a) 
An initial product state evolves under a Hamiltonian, here either an Ising or XXZ model. Correlations develop during dynamics. (b)  In d-NLCE, correlations between two spins separated by $\vec{r}=(x,y)$  already appear in clusters with $N_s=x+y+1$ sites. In contrast, exact diagonalization gives non-zero values of these correlations only by including many more sites,  $N_s=(x+1)(y+1)$.  
}
\label{fig: schematic-NLCE-intuition}
\end{figure}

In this paper, we demonstrate a numerical linked cluster expansion method for calculating dynamics (d-NLCE). We 
show that it dramatically outperforms ED for calculating the short-to-moderate time spin-model dynamics that is common to many ultracold experiments~\cite{vdW_NJP2013,yan_observation_2013,PhysRevLett.113.195302,hazzard:far-from-equilibrium_2013,wallrole,dePaz:nonequilibrium_2013,depaz:probing_2016,
lepoutre:spin_2017,Bohnet1297,Garttner2017,Richerme2014,
zhang:observation_2017,Jurcevic2014,Zeiher2016,orioli:relaxation_2017,
Takei2016,nipper:highly_2012,Jau2016,bernien:probing_2017,
PhysRevLett.115.260401,brown:2D_2015,PhysRevB.96.014303,PhysRevLett.116.247202,heidrich-meisner:quantum_2009,white:correlations_2016,PhysRevLett.117.190602}, illustrated in Fig.~\ref{fig: schematic-NLCE-intuition}(a). It achieves these results with a significantly reduced computational cost.

As we will see, d-NLCE reduces to performing ED on a number of sub-clusters of the full lattice. Remarkably, it often yields accurate results including only a handful of small $N_s\lsim 5$ site clusters. 
d-NLCE results for $N_s\sim 5$ sites are often substantially more accurate than ED with as many as $N_s \sim 10$--$15$ sites. 
This difference in $N_s$ is more striking when recast in terms of the computational cost of the algorithms, since this cost scales exponentially with $N_s$: d-NLCE  is frequently more accurate than ED calculations that require $\sim 10^5$--$10^{10}$ more computational time and memory for the spin-1/2 models that we consider in this paper. 
This advantage
 is expected to be even larger for fermionic or bosonic lattice models, where the Hilbert space grows faster with $N_s$. 

\paragraph{NLCE.} 
NLCEs were introduced in Refs.~\cite{PhysRevLett.97.187202,PhysRevE.75.061118,PhysRevE.75.061119} to calculate observables in equilibrium, as reviewed in Ref.~\cite{TANG2013557}. They have since been used to calculate equilibrium properties of numerous  many-body systems~\cite{rigol:thermodynamic_2009,tang:thermodynamics_2015,khatami:thermodynamics_2011,sahoo:unusual_2016,bruognolo:matrix_2017,benton:quantum_2017,singh:corrections_2012,applegate:vindication_2012,
hayre:thermodynamic_2013,rigol:magnetic_2007,rigol:kagome_2007,khatami:three-dimensional_2016,khatami:finite_2015,khatami:linked_2014,tang:short-range_2012,tang:finite_2013,khatami:effect_2012}, including   entanglement~\cite{sherman:nonzero_2016,sierens:cubic_2017,kallin:corner_2014,
stoudenmire:corner_2014,kallin:entanglement_2013,bruognolo:matrix_2017} and spectra~\cite{sherman:nuclear_2016}, and to investigate many-body localization~\cite{tang:quantum_2015,devakul:early_2015}. More recently they have been used to calculate steady states after a quench~\cite{mallaya:numerical_2017,piroli:correlations_2017,
rigol:fundamental_2016,iyer:optimization_2015,
rigol:quantum_2014,wouters:quenching_2014} and in driven-dissipative systems~\cite{2017arXiv170808666B}. In equilibrium,  NLCEs are the methods of choice for ultracold fermionic atoms in optical lattices at strong interactions and current temperatures~\cite{Cheuk1260,khatami:three-dimensional_2016,PhysRevLett.114.070403,Hart2015a,brown:observation_2016}, complementing determinantal quantum Monte Carlo~\cite{PhysRevD.24.2278} at weak interactions. 
Our work expands the NLCE method to dynamics, and shows that it is especially suitable to the types of dynamics that are now common in experiments in ultracold matter.

We now review the general algorithm for a linked cluster expansion, and describe d-NLCE in particular.
All linked cluster expansions approximate an observable on a graph (such as a lattice) as a summation of this observable on finite connected (linked) clusters of sites. 
The relevant clusters may be any subset of the graph, including the full graph itself.
Define $\expec{A}_c$ as the expectation value of the observable $A$ for the system restricted to the cluster $c$.
Define the weight $W_{A}(c)$
recursively as
\begin{equation}
W_{A}(c) =\expec{A}_c-\underset{s\subset c}{\sum}W_{A}(s) \label{eq: weights}
\end{equation}
where $s\subset c$ indexes every proper subcluster of $c$, that is, every cluster whose components are contained in $c$ except $c$ itself~\footnote{There is considerable flexibility in defining which graphs constitute a cluster}.
From this it follows that
\begin{equation}
\expec{A}_c = \underset{s\subseteq c}{\sum}W_{A}(s) \label{eq: propertyEq}
\end{equation}
where $s\subseteq c$ indexes every subcluster of $c$ (including $c$).
The significance of these definitions is that $W_{A}(c)$  measures only the contribution to $\expec{A}_c$ that arises due to correlations that span the cluster $c$, that is, those where no site is uncorrelated from the rest. 
Thus only linked (i.e. contiguous) clusters need to be considered in Eq.~\eqref{eq: propertyEq}, because 
any unlinked clusters have zero weight. 

We approximate $\expec{A}$ on the full lattice by truncating Eq.~\eqref{eq: propertyEq} to clusters up to size $n$, giving
\begin{eqnarray}
\langle A\rangle^{(n)} &=& \underset{|s|\leq n}{\sum}W_{A}(s)\label{eq: approxOrder}
\end{eqnarray}
where $|s|\leq n$ indexes all the linked subclusters $s$ of the full lattice with number of sites $N_s$ less than or equal to $n$. 
As $n\to \infty$,  
$\langle A\rangle^{(n)} \to \langle A\rangle$.
This approximation converges rapidly as long as correlations decay sufficiently quickly with distance, since in this case $W_{A}(s)$ approaches zero for  sufficiently large clusters $s$.

We see that linked cluster methods contain two steps. First, enumerate linked clusters $s$ up to a certain size (or some other criterion for truncation, such as number of bonds. In this paper, we always use number of sites). Although the computational cost of cluster enumeration grows combinatorically with size, this step has to be performed only once for a given lattice geometry (independent of the Hamiltonian).
Second, solve $\expec{A}_s$ for each cluster and use these to calculate $W_A(s)$.  
Since the clusters are diagonalized independently, NLCE can trivially utilize parallel computing architectures.

Perhaps the most familiar linked cluster method is a high-temperature series expansion (HTSE)~\cite{oitmaa:series_2006}, in which one enumerates all clusters with at most $m$ bonds and solves $\expec{A}_s$ perturbatively to $O(\beta^m)$. 
NLCE replaces the perturbative solution with full diagonalization on each cluster. This is computationally more expensive for each cluster relative to HTSE, but improves the convergence properties for a given expansion order. 

\paragraph{NLCE for dynamics.} Our d-NLCE method applies the NLCE to non-equilibrium systems using the same two steps described above: enumerate the clusters as we would in equilibrium, and calculate $\expec{A}_s(t)$ on each cluster $s$ using full diagonalization.

Figure~\ref{fig: schematic-NLCE-intuition} depicts the dynamics we consider, and suggests a qualitative advantage of d-NLCE over ED for this dynamics. 
The system begins in an initial product state where all spins are aligned in some direction. Then, it evolves under a transverse Ising or XXZ model, as shown in Fig.~\ref{fig: schematic-NLCE-intuition}(a). 
The initial state is uncorrelated, but over time correlations grow between increasingly distant sites. To capture correlations between sites separated by $\xi$ with ED, one must use a system whose linear size is at least $\sim\xi$, as shown in Fig.\ref{fig: schematic-NLCE-intuition}(b). This generically requires a number of sites $N_s\sim \xi^d$ where $d$ is the spatial dimension.
In contrast,  
Fig.~\ref{fig: schematic-NLCE-intuition}(c)  shows how d-NLCE begins to capture these correlations  as soon as one includes clusters with  $N_s \sim \xi$.

We compare d-NLCE with ED for dynamics governed by the transverse Ising  and XXZ models,
\begin{align}
H_{\text{Ising}} &= -J\underset{\left\langle ij\right\rangle}{\sum}\sigma_{i}^{z}\sigma_{j}^{z} - h\underset{i}{\sum}\sigma_{i}^{x},  \label{eq: Ising H} \\
H_{\text{XXZ}} &= -J^{\perp}\underset{\left\langle ij\right\rangle}{\sum}\left(\sigma_{i}^{x}\sigma_{j}^{x} +\sigma_{i}^{y}\sigma_{j}^{y}\right) - J^{z}\underset{\left\langle ij\right\rangle}{\sum}\sigma_{i}^{z}\sigma_{j}^{z}, \label{eq: XXZ H}
\end{align}
where $J$, $J^\perp$, and $J^z$ are spin-spin interactions, $h$ is the transverse field strength, $\sigma_{i}^{\alpha}$ is the Pauli matrix on site $i$ along direction $\alpha$, and $\left\langle ij\right\rangle$ indicates a summation over nearest neighbors.
We consider the system to initially be in a uniform product state, $\bigotimes_i\left(\cos(\theta/2)\ket{\uparrow}_{i}+\sin(\theta/2)\ket{\downarrow}_{i}\right)$.
These two families of spin models are prototypical models in quantum statistical physics, and their dynamics initiated from such product states are being investigated in numerous ultracold experiments~\cite{vdW_NJP2013,yan_observation_2013,PhysRevLett.113.195302,hazzard:far-from-equilibrium_2013,wallrole,dePaz:nonequilibrium_2013,depaz:probing_2016,
lepoutre:spin_2017,Bohnet1297,Garttner2017,Richerme2014,zhang:observation_2017,Jurcevic2014,Zeiher2016,orioli:relaxation_2017,Takei2016,nipper:highly_2012,Jau2016,bernien:probing_2017}.

\begin{figure}[h!]\centering
\includegraphics[width=0.9\columnwidth]{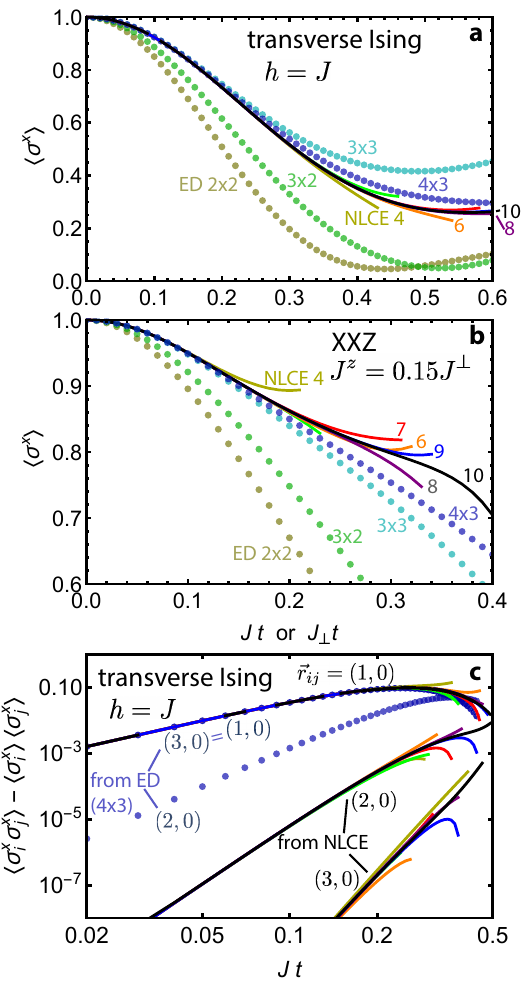}
\caption{
(Color online)  
Comparing d-NLCE and ED for dynamics; method and cluster size are labeled. Magnetization $\braket{\sigma_x^i}$ from initial states   with all spins aligned along $+{\hat x}$, for  (a) the square lattice transverse field Ising model with $h=J$, and
(b) the XXZ model with $J_z=0.15J_x$. 
In both cases, d-NLCE converges accurately for substantially longer post-quench times than ED does, even when d-NLCE uses significantly fewer sites. 
 (c) Connected correlations $\braket{\sigma_i^x\sigma_j^x}-\braket{\sigma_i^x}\braket{\sigma_j^x}$ between sites $i$ and $j$ separated by $\v{r}_{ij}=(1,0)$, $(2,0)$, and $(3,0)$. The curves' color coding in (c) is identical to that in (a) and (b). Even at relatively low $N_s\sim4$, d-NLCE  dramatically outperforms the $N_s=12=4\!\times \!3$ ED 
 for separations of $(2,0)$ and $(3,0)$.
 }
\label{fig: NLCE vs ED}
\end{figure}

Figure~\ref{fig: NLCE vs ED} compares d-NLCE to ED with periodic boundary conditions for several cluster sizes. 
Throughout we choose an initial state with spins polarized along the $\hat x$ direction, $\bigotimes_i \ket{\rightarrow}_i$. Figures~\ref{fig: NLCE vs ED}(a)  and~(b) show  $\langle\sigma^{x}_i\rangle(t)$ for the $h=J$ transverse Ising and $J^{z} = 0.15J^{\perp}$ XXZ models on a two-dimensional square lattice.  
These parameters are chosen to be representative.  
In both ED and d-NLCE, $\braket{\sigma^x_i}(t)$ converges as $N_s$ is increased. 
In this way, both d-NLCE and ED monitor their own convergence by how closely successive $N_s$ results coincide with each other.

While both d-NLCE and ED converge, Fig.~\ref{fig: NLCE vs ED}  shows that d-NLCE remains  accurate to significantly longer times than ED, even when using somewhat smaller clusters. 
For example, for the transverse Ising results shown in Fig.~\ref{fig: NLCE vs ED}(a), d-NLCE has visually converged for $Jt\lsim 0.6$ using $N_s=8$ site clusters. In contrast,  ED shows clear discrepancies at times as short as $Jt \approx 0.25$, even while using significantly larger clusters with $N_s=12=4\times3$, which requires an exponentially greater computational cost. Even the $N_s=4$ d-NLCE is accurate to longer times than the $N_s=12$ ED.
This trend  also holds for the XXZ model: Fig.~\ref{fig: NLCE vs ED}(b) shows that the $N_s=10$ d-NLCE converges for $J^{\perp}t \lsim 0.3$, while the more computationally demanding $N_s=12=4\times 3$ ED has noticeable discrepancies starting at $J^{\perp}t \approx 0.15$.

\begin{figure}\centering
\includegraphics[width=0.9\columnwidth]{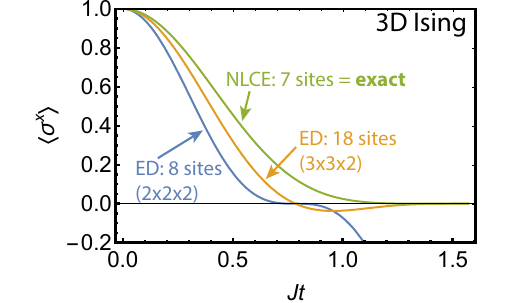}
\caption{(Color online) Dynamics of the magnetization 
$\braket{\sigma_x^i(t)}$ 
in the three-dimensional cubic lattice Ising model obtained from d-NLCE and ED, with no transverse field ($h=0$). This dynamics can be solved exactly, and d-NLCE yields exact results already  at $N_s=7$ sites, while ED has significant deviations from the exact results yield exact results even at $N_s=18=3\!\times\!3\!\times\!2$ sites. 
\label{fig: 3D Ising}}
\end{figure}

d-NLCE also captures the growth of correlations in situations where  ED fails severely. Figure~\ref{fig: NLCE vs ED}(c) shows the dynamics of $\expec{\sigma^x_i\sigma^x_j}-\expec{\sigma^x_i}\expec{\sigma^x_j}$ for sites $i$ and $j$ separated by $(1,0)$, $(2,0)$, and $(3,0)$  for the two-dimensional transverse Ising model.  At short times, each of these grows polynomially with time.  For each separation shown, d-NLCE captures this behavior exactly in any calculation with at least $N_s=4$. In contrast, ED fails dramatically even when it uses a substantially larger $N_s = 12 = 4\times3$. It predicts, for example, that the $(3,0)$ correlation is the same as the $(1,0)$ correlation (due to periodic boundary conditions), a result that is manifestly wrong and inaccurate by several orders of magnitude.  As an aside, we  note that while this leading order behavior can be captured by a  short-time series expansion (STSE), the dynamical analog of the high temperature series expansion,  the leading $t^{10}$ behavior of the $(3,0)$ correlations  would require an STSE involving clusters with up to 10 bonds (up to 11 sites). In contrast, d-NLCE requires just $N_s=4$.

The d-NLCE method compares even more favorably with ED in three dimensions than in the two-dimensional models that we have just considered. 
As a particularly favorable example,  Fig.~\ref{fig: 3D Ising} shows the $h=0$ Ising dynamics of $\expec{\sigma^x_i}(t)$ on a cubic lattice.  
The $N_s=7$ d-NLCE reproduces the exact answer, which can be obtained analytically for this system~\cite{PhysRevA.87.042101,2013NJPh...15k3008F,Kastner_analytic,Radin_analytic}, but ED is not accurate even on an $N_s=18=3\times3\times2$ site system. 
The minimum size required by ED to reproduce the exact result is in fact a $3\times3\times3 = 27$ site cluster with periodic boundaries,
 which requires $\sim 10^{15}$--$10^{18}$ times more computational resources than $N_s=7$ 
d-NLCE~\footnote{This is a crude estimate: ED requires 20 additional sites, and hence a $2^{20}\approx 10^6$ times larger Hilbert space. Matrix diagonalization scales as the cube of this, giving $10^{18}$.  However, d-NLCE requires a handful of clusters (though this is trivially parallelized) and accounting for symmetries decreases the ED Hilbert space, so a conservative factor of $10^{15}$  may be more appropriate.}.   
We expect d-NLCE's increased advantage over ED in three dimensions to hold quite  generally.   This is because, as illustrated in Fig.~\ref{fig: schematic-NLCE-intuition}(b), clusters begin spanning the system's correlation length at smaller $N_s$.   

\begin{figure}\centering
\includegraphics[width=0.98\columnwidth]{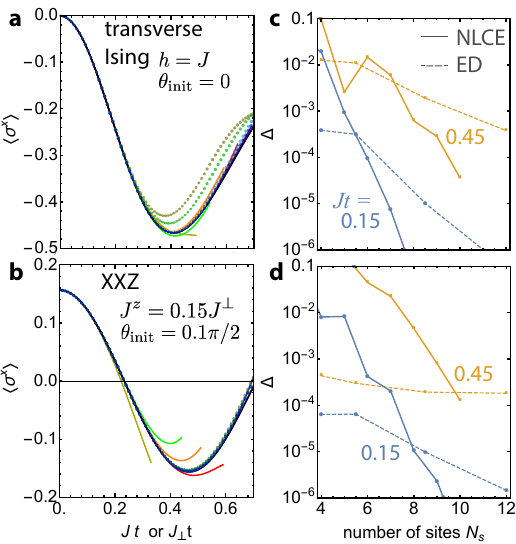}
\caption{  (Color online)  Comparing d-NLCE ($N_s=4,\ldots,10$) and ED ($2\!\times \!2$, $3\!\times \!2$, $3\!\times\! 3$, $4\!\times \!3$) for dynamics 
of $\braket{\sigma_x^i(t)}$ in (a) the $h=J$ transverse field Ising model initiated from a state with spins at $t=0$ aligned  along the $\hat z$ axis, and (b) the $J^\perp=0.15J^z$ XXZ model on a square lattice with spins at $t=0$ aligned $\theta_{\text{init}}=0.1\pi/2$ away from the $\hat z$ axis in the $x$-$z$ plane. 
The curves' color coding is identical to that in Fig.~\ref{fig: NLCE vs ED}.
(c,d) The improvement $\Delta_n(t)$ provided by advancing from $(n-1)$ sites to $n$ sites in d-NLCE (solid) and ED (dashed).  
Thus we see that although both methods work well, at these times d-NLCE is more accurate for less computational cost.
}
\label{fig: NLCE vs ED2}
\end{figure}

Finally, we plot the dynamics of $\left\langle \sigma_x \right\rangle$ in Figs.~\ref{fig: NLCE vs ED2}(a,b) for two cases with different initial conditions from Fig.~\ref{fig: NLCE vs ED}. In these cases, both d-NLCE and ED appear to work well by visual inspection. 
In the transverse Ising model with these initial conditions, d-NLCE converges at most times shown for $N_s=4$ and for all times shown by $N_s=9$, whereas ED has serious discrepancies at most times for $N_s=6=3\times 2$ and is visually converged for $N_s=9=3\times 3$. For the XXZ model, all of the ED results shown appear to be accurate, while d-NLCE  converges only around $N_s=8$.

Although Figs.~\ref{fig: NLCE vs ED2}(a,b) indicate that ED provides comparable results to d-NLCE in some situations, in all the cases we study, d-NLCE converges more quickly as a function of cluster size. 
This is clearly shown in Figs.~\ref{fig: NLCE vs ED2}(c,d), where we plot the difference $\Delta$ between results at two consecutive $N_s$~\footnote{We calculate ED results at $N_s=4, 6, 9$ and $12$. For other $N_s$, we interpolate these results to determine the $\Delta$s } (at a fixed time), as a function of $N_s$.
Both methods converge roughly exponentially. However, d-NLCE converges exponentially faster, i.e. d-NLCE provides a greater increase in accuracy for each increase in $N_s$,  in accord with the intuition presented in Fig.~\ref{fig: schematic-NLCE-intuition}(b). The d-NLCE's $\Delta$ becomes smaller than that of ED for $N_s$ between 5 and 10.

\paragraph{Conclusions.} We have developed a numerical linked cluster expansion for out-of-equilibrium quantum dynamics and demonstrated that it is able to accurately capture dynamics relevant to numerous ongoing experiments in ultracold matter. 
We compared d-NLCE against ED, often the best previously available method, and found that d-NLCE (i)~converges to longer times while utilizing  the same number of sites, (ii)~converges more quickly as a function of the cluster size, and (iii)~captures leading-order evolution of correlations with a drastically reduced cluster size. 
The reduced cluster sizes required result in exponentially faster computation. 
The  advantages of d-NLCE were clear in two-dimensions, and even larger in three dimensions. They are expected to be more pronounced for systems with larger on-site Hilbert spaces, like large spins or Fermi-Hubbard models, where the Hilbert space dimension grows even faster with cluster size, compounding d-NLCE's advantage of utilizing smaller clusters than ED for a given level of accuracy.

The basic d-NLCE method used here can be improved and optimized in several ways. 
Although we sum all clusters up to a maximum size, choosing different sets of clusters can improve accuracy. 
Furthermore, equilibrium NLCE computations benefit from resummation techniques, and we expect that applying resummation to d-NLCE will improve its convergence as well, although different resummation techniques may be necessary out of equilibrium. 
Finally, calculations can be extended to higher order by improvements in the implementation: for example, numerically solving the cluster dynamics using Krylov subspace methods~\cite{schmitteckert:nonequilibrium_2004,garcia-ripoll:time_2006}, perhaps in combination with tensor network methods.  d-NLCEs including clusters to $N_s=20$ or more sites are likely feasible with such improvements.

\begin{acknowledgments} 
This material is based upon work supported with funds
from the Welch Foundation,  grant no.  C-1872.  KRAH
thanks the Aspen Center for Physics, which is supported
by the National Science Foundation grant PHY-1066293,
for its hospitality while part of this work was performed. 
\end{acknowledgments}

\bibliography{dNLCE-v2}

\end{document}